\documentclass[pra,twocolumn,amsmath,amssymb,amsfonts,showkeys,showpacs,preprintnumbers]{revtex4}
\usepackage{bm,graphicx,xcolor,subfigure,hyperref}
\newcommand{\eps}{\varepsilon}
\newcommand{\ml}{\mathcal{L}}
\newcommand{\mc}{\multicolumn}
\DeclareMathOperator{\Li}{Li}
\DeclareMathOperator{\Lag}{L}
\begin{document}
\bibliographystyle{apsrev}

	\title{Ground state of two electrons on a sphere}
	
	\author{Pierre-Fran\c{c}ois Loos}
	\affiliation{Research School of Chemistry, Australian National University, Canberra ACT 0200, Australia}

	\author{Peter M. W. Gill}
 	\thanks{Corresponding author}
	\email{peter.gill@anu.edu.au}
	\affiliation{Research School of Chemistry, Australian National University, Canberra ACT 0200, Australia}

	\date{\today}

	\begin{abstract}
	We have performed a comprehensive study of the singlet ground state of two electrons on the surface of a sphere of radius $R$.  We have used electronic structure models ranging from restricted and unrestricted Hartree-Fock theory to explicitly correlated treatments, the last of which lead to near-exact wavefunctions and energies for any value of $R$.  M{\o}ller-Plesset energy corrections (up to fifth-order) are also considered, as well as the asymptotic solution in the large-$R$ regime.
	\end{abstract}

	\maketitle
\section{\label{sec:intro}Introduction}

Exactly (or very accurately) solvable models have ongoing value and are valuable both for illuminating more complicated systems and for testing theoretical approaches, such as density functional methods \cite{HohenbergPRB1964, KohnPRA1965, ParrYang}.  One such model is the Hooke's law atom (or Harmonium) which is composed of two electrons bound to a nucleus by a harmonic potential but repelling Coulombically.  This system was first considered more than 40 years ago by Kestner and Sinanoglu \cite{KestnerPhysRev1962}, but solved analytically in 1989 by Kais {\em et al.} \cite{KaisJCP1989} for a particular value of the harmonic force constant and, later, for a countably infinite set force constants \cite{TautPRA1993}.

A related system, studied by Alavi and co-workers \cite{AlaviJCP2000,ThompsonPRB2002,ThompsonJCP2005}, consists of two electrons, interacting through a Coulomb potential, but confined within a ball of radius $R$.  This possesses a number of interesting features, including the formation of a ``Wigner molecule'' for large $R$ \cite{ThompsonPRB2004}.  The spontaneous formation of such molecules can also occur in quantum dots and is analogous to the Wigner crystallization \cite{WignerPR1934} of the uniform electron gas.

If the two electrons are constrained to remain on the \emph{surface} of the sphere, one obtains a model that Berry and co-workers have used \cite{EzraPRA1982, EzraPRA1983, OjhaPRA1987, HindePRA1990} to understand both weakly and strongly correlated systems, such as the ground and excited states of the helium atom, and also to suggest the ``alternating'' version of Hund's rule \cite{WarnerNature1985}.  Seidl studied this system in the context of density functional theory \cite{SeidlPRA2007b} in order to test the ISI (interaction-strength interpolation) model \cite{SeidlPRL2000}.  For this purpose, he derived accurate solutions in both the weak interaction limit (the small $R$ regime) and the strong interaction limit (the large $R$ regime).  He also obtained accurate results by numerical integration of the Schr{\"o}dinger equation.

In this paper, we are interested in the $^1S$ ground state of two electrons on the surface of a sphere of radius $R$.  This allows us to restrict our study to the symmetric spatial part of the wavefunction and ignore the spin coordinates.  We have extended Seidl's analysis and performed an exhaustive study using a range of models. We restrict our analysis to the repulsive potential case; the strong-attraction limit (attractive potential) is carefully examined in Ref. \cite{SeidlPRA2007b}.

Restricted and unrestricted Hartree-Fock (HF) solutions are discussed in Section \ref{sec:HF} and the strengths and weaknesses of M{\o}ller-Plesset (MP) perturbation theory \cite{MollerPhysRev1934} in Section \ref{sec:MP}.  We consider asymptotic solutions for large $R$ in Section \ref{sec:LargeR} and, in Section \ref{sec:R12}, we explore several variational schemes including explicitly correlated techniques \cite{HylleraasAdvQuantumChem1964, KutzelniggTheorChemAcc1985, KlopperCPL1987, KlopperJPhysChem1990, KutzelniggJChemPhys1991}) that enforce the cusp condition \cite{Kato1957, PackJChemPhys1966}. Atomic units are used throughout.

\section{\label{sec:hamiltonian}Hamiltonian}

The absolute position of the $i$-th electron is defined by its spherical polar angles $\bm\Omega_i = \left( \theta_i, \phi_i \right)$. 
The relative position of the electrons is conveniently measured by the interelectronic angle $\theta$ which they subtend at the origin. 
These coordinates are related by
\begin{equation}
	\cos \theta = \cos \theta_1 \cos \theta_2 + \sin \theta_1 \sin \theta_2 \cos \left( \phi_1 - \phi_2 \right) 
\end{equation}
and we have $0 \le u \equiv | \bm{r}_1 - \bm{r}_2 | \le 2R$.

The Hamiltonian is
\begin{equation} \label{H}
	\hat{H} = \hat{T} + u^{-1}
\end{equation}
where 
\begin{equation} \label{T}
	\hat{T} = \hat{T}_1 + \hat{T}_2 = - \frac{\nabla_1^2}{2} - \frac{\nabla_2^2}{2}
\end{equation}
is the kinetic operator for both electrons and $u^{-1}$ is the Coulomb operator.  In terms of $\theta$, the Hamiltonian is
\begin{equation} \label{H-theta}
	\hat{H} = - \frac{1}{R^2} \left( \frac{d^2}{d \theta^2} + \cot \theta \frac{d}{d \theta} \right) + \frac{1}{2 R} \csc \frac{\theta}{2}
\end{equation}
in which form it becomes clear that the kinetic and potential parts of $\hat{H}$ scale with $R^{-2}$ and $R^{-1}$, respectively.

\section{\label{sec:HF}Hartree-Fock approximations}

\subsection{\label{subsec:RHF}Restricted Hartree-Fock}

In the HF approximation, each electron feels the mean field generated by the other electron \cite{Szabo}.  The restricted Hartree-Fock (RHF) solution
\begin{equation}
	\Phi^{\rm RHF} (\bm\Omega_1,\bm\Omega_2) = \Psi^{\rm RHF}(\bm\Omega_1)\,\Psi^{\rm RHF}(\bm\Omega_2)
\end{equation}
places both electrons in an orbital $\Psi^{\rm RHF}$ that is an eigenfunction of the Fock operator
\begin{equation} \label{F}
	\hat{F}_1= \hat{T}_1 + \int \frac{\left|\Psi^{\rm RHF}(\bm{\Omega}_2)\right|^2}{u}\,R^2\,d\bm\Omega_2
\end{equation}
with $d\bm\Omega_2 = \sin \theta_2\,d\theta_2\,d\phi_2$.

By definition, the one-electron basis function
\begin{equation}
	\Psi_{\ell m} (\bm\Omega_i) = \frac{Y_{\ell m} (\bm\Omega_i)}{R}
\end{equation}
where $Y_{\ell m}$ is the spherical harmonic of degree $\ell$ and order $m$ is an eigenfunction of $\Hat{T}_i$ with eigenvalue $\ell(\ell+1)/(2R^2)$. 
Using the partial-wave expansion \cite{Arfken}
\begin{equation} \label{coulomb-expansion}
	u^{-1} = R^{-1} \sum_{\ell=0}^\infty P_\ell(\cos\theta)
\end{equation}
and the addition theorem \cite{Abramowitz}
\begin{equation}
	P_{\ell} (\cos\theta) = \frac{4 \pi}{2\ell+1} \sum_{m=-\ell}^{+\ell} Y_{\ell m}^{\star} (\bm\Omega_1)\,Y_{\ell m} (\bm\Omega_2)
\end{equation}
it is straightforward to show that 
\begin{equation}
	\int \frac{|\Psi_{00} (\bm\Omega_2)|^2}{u}\,R^2\,d\bm\Omega_2 = \frac{1}{R}
\end{equation}
The orbital $\Psi_{00} (\bm\Omega_i)$ is thus an eigenfunction of $\hat{F}_i$ with the eigenvalue $1/R$. 
Moreover, it follows from the orthogonality of the spherical harmonics that
\begin{equation}
	\left< \Psi_{\ell m} (\bm\Omega_1) \left| \Hat{F}_1 \right| \Psi_{00} (\bm\Omega_1) \right> = \delta_{\ell,0}\,\delta_{m,0}
\end{equation}
which ensures the stationarity of the RHF energy with respect to the orbitals $\Psi_{\ell m}$.

The ground-state RHF energy is thus
\begin{equation} \label{ERHF}
	E^{\rm RHF} = \frac{1}{R}
\end{equation}
and the normalized RHF wavefunction is
\begin{equation} \label{phiRHF}
	\Phi^{\rm RHF} = \frac{1}{4 \pi R^2}
\end{equation}
which yields a uniform electron density over the surface of the sphere.

\subsection{\label{subsec:UHF}Unrestricted Hartree-Fock}

When $R$ exceeds a critical value, a second, unrestricted HF (UHF) solution develops \cite{CizekJCP1967, CizekJCP1970, SeegerJCP1977} in which the two electrons tend to localize on opposite sides of the sphere.  This is analogous to the UHF description of a dissociating H$_2$ molecule \cite{Szabo}.

To obtain this symmetry-broken solution
\begin{equation}\label{Psi-UHF}
	\Phi^{\rm UHF} (\theta_1,\theta_2) = \Psi^{\rm UHF} (\theta_1)\,\Psi^{\rm UHF} (\pi - \theta_2)
\end{equation}
we expand the orbital as
\begin{equation}\label{basis-UHF}
	\Psi^{\rm UHF} (\theta_i) = \sum_{\ell=0}^{\infty} C_{\ell} \,\Psi_{\ell} (\theta_i)
\end{equation}
where the $\Psi_{\ell} (\theta_i) = Y_{\ell} (\theta_i)/R = Y_{\ell 0} (\bm\Omega_i)/R$ are zonal spherical harmonics. 
The Fock matrix elements in this basis are
\begin{align} \label{F-UHF}
	F_{\ell_1 \ell_2}& = \left< \Psi_{\ell_1} (\theta_i) \left| \Hat{F}_i \right| \Psi_{\ell_2} (\theta_i) \right>	\nonumber	\\
					& = \frac{\ell_1(\ell_1+1)}{2 R^2} \delta_{\ell_1,\ell_2}
						+ \sum_{\ell_3,\ell_4=0}^\infty C_{\ell_3} C_{\ell_4} G_{\ell_1 \ell_2}^{\ell_3 \ell_4}
\end{align}
where the two-electron integrals are
\begin{equation}\label{G-UHF}
	G_{\ell_1 \ell_2}^{\ell_3 \ell_4} = \left< \Psi_{\ell_1} (\theta_1)\,\Psi_{\ell_3} (\theta_2) \left| u^{-1} \right| \Psi_{\ell_2} (\theta_1)\,\Psi_{\ell_4} (\theta_2) \right>
\end{equation}
Using the partial-wave expansion \eqref{coulomb-expansion} and the relation
\begin{align} \label{Y-W3J}
	\left< \ell_1\,\ell_2\,\ell_3 \right>	& = \int Y_{\ell_1} (\theta)\,Y_{\ell_2} (\theta)\,Y_{\ell_3} (\theta) \sin\theta\,d\theta	\nonumber	\\
						& = \sqrt{\frac{(2\ell_1+1)(2\ell_2+1)(2\ell_3+1)}{4 \pi}}	
							\begin{pmatrix}
								\ell_1	&	\ell_2	&	\ell_3	\\
									0	&		0	&		0	\\
							\end{pmatrix}^2
\end{align}
between the integrals of three spherical harmonics and the Wigner 3j-symbols \cite{Edmonds}, we find
\begin{equation}\label{G}
	G_{\ell_1 \ell_2}^{\ell_3 \ell_4} = \frac{(-1)^{\ell_3+\ell_4}}{R}
		\sum_{\ell=0}^{\infty} \frac{4 \pi}{2\ell+1} \left< \ell_1\,\ell_2\,\ell \right> \left< \ell_3\,\ell_4\,\ell \right>
\end{equation}
where selection rules \cite{Edmonds} restrict the terms in the sum.

The UHF energy is then given by
\begin{equation}\label{E-UHF}
	E^{\rm UHF} = \sum_{\ell=0}^{\infty} C_{\ell}^2 \frac{\ell(\ell+1)}{R^2} + \sum_{\ell_1,\ell_2=0}^{\infty} C_{\ell_1} C_{\ell_2} F_{\ell_1 \ell_2}
\end{equation}
The first term is the kinetic energy and is positive.  However, for sufficiently large $R$, this is outweighed by negative contributions in the second term and it is these that drive the symmetry-breaking process.

For computational reasons, we truncate the sum in \eqref{basis-UHF} at $\ell = L$ but, for all of the radii $R$ considered in this study, we found that $L = 15$ suffices to obtain $E^{\rm UHF}$ with an accuracy of $10^{-12}$.

\begin{center}
\begin{figure}
\caption{\label{fig:SBHF}$R \times E^{\rm RHF}$ (dashed), $R \times E^{\rm UHF}$ (dotted) and $R \times E^{\rm exact}$ (solid) as a function of $R$.}
	\includegraphics[width=0.48\textwidth]{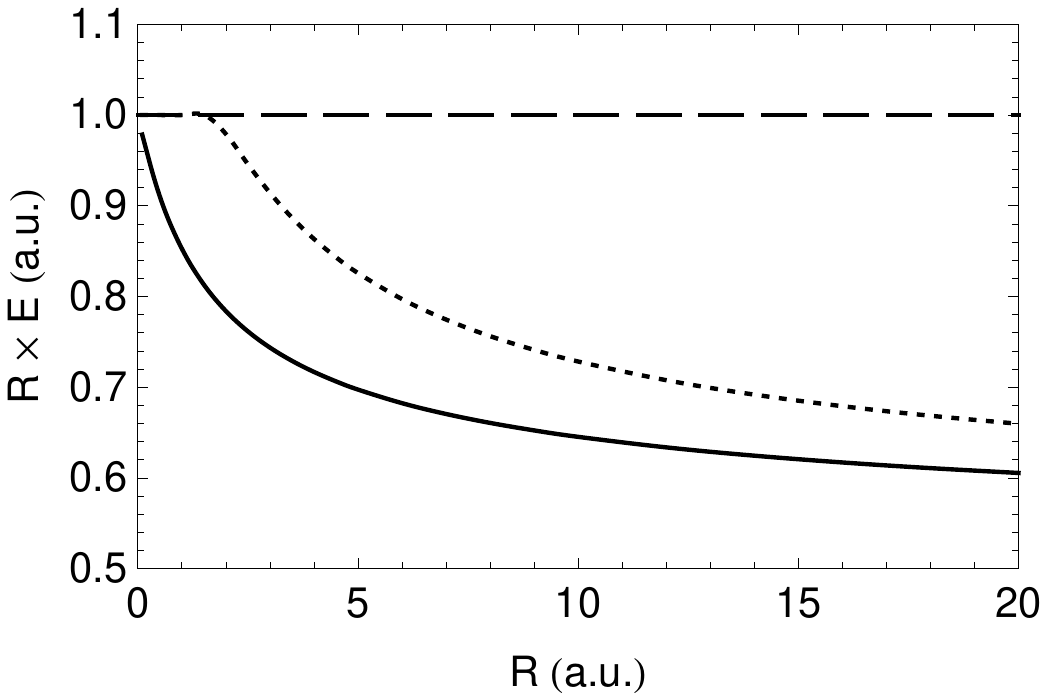}
\end{figure}
\end{center}

As Table \ref{tab:SBHF} and Figure \ref{fig:SBHF} show, the UHF solution becomes lower than the RHF one for $R > R^{\rm crit} \approx 1.5$ and the UHF, not RHF, energy behaves correctly for large $R$.  Specifically, it can be shown that
\begin{gather}
	\lim_{R \to \infty} R\,E^{\rm RHF} = 1\\
	\lim_{R \to \infty} R\,E^{\rm UHF} = 1/2
\end{gather}
The UHF result reflects the Coulomb interaction between two electrons localized on opposite sides of the sphere \cite{SeidlPRA2007b}, a phenomenon known as Wigner crystallization \cite{ThompsonPRB2004, WignerPR1934}.  The difference between the UHF and exact energies (\textit{i.e.} the correlation energy) appears to decay as $O(R^{-3/2})$.

\begin{table}
\caption{\label{tab:SBHF}RHF, UHF and exact energies for various $R$.}
\begin{ruledtabular}
\begin{tabular}{lllr}
\mc{1}{c}{$R$}	&	\mc{1}{c}{$E^{\rm RHF}$}	&	\mc{1}{c}{$E^{\rm UHF}$}	&	\mc{1}{c}{$E^{\rm exact}$}	\\
\hline
0.0001			&	10000						&	10000						&	9999.772 600 495 			\\
0.001			&	1000						&	1000						&	  999.772 706 409			\\
0.01			&	100							&	100							&	    99.773 761 078			\\
0.1				&	10							&	10							&		9.783 873 673			\\
0.2				&	5							&	5							&		4.794 237 154			\\
0.5				&	2							&	2							&		1.820 600 768			\\
1				&	1							&	1							&		0.852 781 065			\\
2				&	0.500 000					&	0.489 551					&		0.391 958 796			\\
3				&	0.333 333					&	0.304 783					&		0.247 897 526			\\
4				&	0.250 000					&	0.215 864					&		0.179 210 308			\\
5				&	0.200 000					&	0.165 161					&		0.139 470 826			\\
10				&	0.100 000					&	0.072 829					&		0.064 525 123			\\
20				&	0.050 000					&	0.032 983					&		0.030 271 992			\\
50				&	0.020 000					&	0.012 006					&		0.011 363 694			\\
100				&	0.010 000					&	0.005 708 105				&		0.005 487 412			\\
1000			&	0.001 000					&	0.000 522 363				&		0.000 515 686			\\
\end{tabular}
\end{ruledtabular}
\end{table}

\section{\label{sec:MP}Expansion for small R}

In M{\o}ller-Plesset perturbation theory, the Hamiltonian of the system is partitioned as
\begin{equation}
	\Hat{H} = \Hat{H}_0 + \Hat{V}
\end{equation}
where $\Hat{H}_0$ the zeroth-order Hamiltonian and $\Hat{V}$ is a perturbation operator and, in our case, we have
\begin{align}
	\Hat{H}_0	& = \Hat{T}	\\
	\Hat{V}		& = u^{-1}
\end{align}
The ground-state wavefunction and energy are expanded
\begin{gather}
	 \Phi = \Phi^{(0)} + \Phi^{(1)} + \Phi^{(2)} + \Phi^{(3)} + \ldots\\
	 E = E^{(0)} + E^{(1)} + E^{(2)} + E^{(3)} + \ldots
\end{gather}
We will refer to $E^{(n)}$ as the $n$th-order energy and define the MP$n$ correlation energy as
\begin{equation}
	E^{\text{MP}n} = \sum_{m=2}^n E^{(m)}
\end{equation}
Dimensional analysis reveals that
\begin{gather}
	\Phi	= \frac{\phi_0}{R^2} + \frac{\phi_1}{R} + \phi_2 + \phi_3 R + \phi_4 R^2 + \phi_5 R^3 + \ldots		\label{expansion-Psi}	\\
	E	= \frac{\eps_0}{R^2} + \frac{\eps_1}{R} + \eps_2 + \eps_3 R + \eps_4 R^2 + \eps_5 R^3 + \ldots	\label{expansion-E}
\end{gather}
where the $\phi_n$ are functions of $\theta$ and the $\eps_n$ are numbers. 
From \eqref{ERHF} and \eqref{phiRHF}, we see $\phi_0 = 1/4\pi$, $\eps_0 = 0$ and $\eps_1 = 1$.

The excited eigenfunctions of $\Hat{H}_0$ are given by
\begin{equation}
	\Phi_{\ell_1 m_1}^{\ell_2 m_2} (\bm\Omega_1,\bm\Omega_2) = \Psi_{\ell_1 m_1} (\bm\Omega_1)\,\Psi_{\ell_2 m_2} (\bm\Omega_2)
\end{equation}
and we can expand the exact wavefunction $\Phi$ in this basis. However, for the $^1S$ ground state, angular momentum theory \cite{Edmonds,Slater} 
limits the combinations of $\ell_1$, $\ell_2$, $m_1$ and $m_2$ that contribute and it is more efficient to expand $\Phi$ in the basis of two-electron functions
\begin{equation}\label{L}
	\Phi_\ell(\theta) = \frac{\sqrt{2\ell+1}}{4 \pi R^2}\ P_\ell( \cos \theta )
\end{equation}
which are eigenfunctions of $\hat{T}$ with eigenvalues
\begin{equation}
	E_\ell = \frac{\ell(\ell+1)}{R^2}
\end{equation}

\subsection{First-order wavefunction}

In the intermediate normalization, the first-order wavefunction is
\begin{align} \label{MP1}
	\Phi^{(1)}(\theta) \equiv \frac{\phi_1(\theta)}{R}
	& = \sum_{\ell=1}^\infty \frac{\left< \Phi^{\rm RHF} \left| \hat{V} \right| \Phi_\ell \right>}{E_0 - E_\ell}\ \Phi_\ell(\theta)	\nonumber	\\
	& = - \frac{1}{4\pi R} \sum_{\ell=1}^\infty \frac{1}{\ell(\ell+1)}\ P_\ell(\cos\theta)
\end{align}
Using the Legendre generating function
\begin{equation}\label{gen-Legendre}
	\sum_{\ell=0}^\infty P_\ell(x)\,t^\ell = \frac{1}{\sqrt{1 - 2\,x\,t + t^2}}
\end{equation}
the sum in \eqref{MP1} can be found in closed-form, yielding
\begin{equation}
	\Phi^{(1)}(\theta) = \frac{1}{4 \pi R} \left[ 2 \ln \left( 1 + \sin \frac{\theta}{2} \right) - 1 \right]
\end{equation}
or, equivalently,
\begin{equation}
	\Phi^{(1)}(u) = \frac{1}{4\pi R} \left[ 2\ln \left( 1 + \frac{u}{2R} \right) - 1 \right]
\end{equation}
and these yield the normalized first-order wavefunction
\begin{equation} \label{Phi-MP1}
	\Phi^{\rm MP1}(\theta) = \frac{ \Phi^{\rm RHF} + \Phi^{(1)} (\theta) }{\sqrt{1 + (16\ln2 - 11)R^2}}
\end{equation}

The true ground-state wavefunction must be nodeless. However, it is easy to show that the MP1 wavefunction possesses a node if $R > 1$, leading us to anticipate that $\Phi^{\rm MP1}$ will be a poor wavefunction for large spheres.

\subsection{Second- and third-order energies}

According to the Wigner $2n+1$ rule \cite{Helgaker}, the 1st-order wavefunction generates the 2nd- and 3rd-order energies.  The 2nd-order energy, which has previously been found by Seidl \cite{SeidlPRA2007b}, is given by
\begin{align} \label{EMP2}
	E^{(2)} \equiv \eps_2	& = \left< \Phi^{\rm RHF} \left| \hat{V} \right| \Phi^{(1)} \right>	\nonumber	\\
							& = 4 \ml - 3													\nonumber	\\
							& = -0.227\;411\;278\ldots
\end{align}
where $\ml = \ln2$.  As Table \ref{tab:Ecorr} shows, the MP2 correlation energy is an excellent approximation for small $R$ but, because it is independent of $R$, it is poor for large $R$.

It is surprising to find that $E^{(2)}$ is so much larger than the limiting correlation energies \cite{GillJCP2005} of the helium-like ions ($-0.0467$) or Hooke's Law atoms ($-0.0497$).

The 3rd-order energy is given by
\begin{equation}
	E^{(3)} \equiv \eps_3 R = \left< \Phi^{(1)} \left| \hat{V} - E^{(1)} \right| \Phi^{(1)} \right>
\end{equation}
and this yields
\begin{align} \label{EMP3}
	\eps_3	& = 8 (\ml^2 - 5\ml + 3)	\nonumber	\\
			& = +0.117\;736\;889\ldots
\end{align}
which agrees with Seidl's rough estimate \cite{SeidlPRA2007b}.  Table \ref{tab:Ecorr} shows that MP3 gives an improvement over MP2 but that it, too, eventually breaks down as $R$ increases.

\begin{turnpage}
\begin{table*}
\caption{\label{tab:Ecorr}Correlation energies (relative to UHF and multiplied by $-1$) from various models for various $R$.}
\begin{ruledtabular}
\begin{tabular}{ccccccccccc}
R 			&MP2		&MP3				&MP4				&MP5				&Hylleraas		&Seidl				&Exact\footnotemark[1]\\
\hline
0.0001			&{0.227} 411	&{0.227 399 504} 071		&{0.227 399 504 574}		&{0.227 399 504 574}			&{~0.22}2 212	&---				&0.227 399 504 574\\
0.001\phantom{0}	&{0.227} 411	&{0.227 293 5}41\phantom{ 000}	&{0.227 293 591 1}47		&{0.227 293 591 133}			&{~0.22}2 123	&---				&0.227 293 591 133\\
0.01\phantom{00}	&{0.22}7 411	&{0.226 23}4\phantom{ 000 000}	&{0.226 238 9}36\phantom{ 000}	&{0.226 238 922 4}73			&{~0.22}1 237	&---				&0.226 238 922 463\\
0.1\phantom{000}	&{0.2}27 411	&{0.21}5 638\phantom{ 000 000}	&{0.216 1}40\phantom{ 000 000}	&{0.216 126 3}87\phantom{ 000}		&{~0.21}2 574	&0.2175\footnotemark[2]	&0.216 126 326 630\\
0.2\phantom{000}	&{0.2}27 411	&{0.20}3 864\phantom{ 000 000}	&{0.205} 875\phantom{ 000 000}	&{0.205 76}3 261\phantom{ 000}		&{~0.20}3 406	&0.2064\footnotemark[2]	&0.205 762 846 030\\
0.5\phantom{000}	&{0}.227 411	&{0.1}68 543\phantom{ 000 000}	&{0.1}81 112\phantom{ 000 000}	&{0.179 3}67\phantom{ 000 000}		&{~0.17}8 908	&0.1796\footnotemark[2]	&0.179 399 232 168\\
1\phantom{.0000}	&{0}.227 411	&{0.1}09 674\phantom{ 000 000}	&{0.1}59 950\phantom{ 000 000}	&{0.14}5 992\phantom{ 000 000}		&{~0.147} 181	&0.1473\footnotemark[2]	&0.147 218 934 944\\
\cline{1-5}                                                                                                                             
			&$e-e$		&LR0				&LR1			&LR2\\                                  
\cline{1-5}                                                                                                                             
2\phantom{.0000}	&{0}.239 551	&{0.0}62 774\phantom{ 000}	&{0.09}4 024\phantom{ 000 000}	&{0.09}5 405\phantom{ 000 000}		&{~0.09}6 444	&0.0977\footnotemark[3]	&0.097 591 955 594\\
3\phantom{.0000}	&{0}.138 116	&{0.0}41 890\phantom{ 000}	&{0.05}5 780\phantom{ 000 000}	&{0.056} 281\phantom{ 000 000}		&{~0.05}4 783	&---				&0.056 885 070 442\\
4\phantom{.0000}	&{0.0}90 864	&{0.0}28 352\phantom{ 000}	&{0.036} 176\phantom{ 000 000}	&{0.036} 420\phantom{ 000 000}		&{~0.03}3 984	&---				&0.036 653 426 934\\
5\phantom{.0000}	&{0.0}65 161	&{0.02}0 440\phantom{ 000}	&{0.025} 440\phantom{ 000 000}	&{0.025} 580\phantom{ 000 000}		&{~0.02}2 707	&0.0257\footnotemark[3]	&0.025 690 364 031\\
10\phantom{.000}	&{0.0}22 829	&{0.00}7 018\phantom{ 000}	&{0.008} 268\phantom{ 000 000}	&{0.008} 292\phantom{ 000 000}		&{~0.00}5 129	&0.0083\footnotemark[3]	&0.008 303 955 973\\
20\phantom{.000}	&{0.00}7 983	&{0.002} 393\phantom{ 000}	&{0.002 7}06\phantom{ 000 000}	&{0.002 71}0\phantom{ 000 000}		&{~0.00}0 154	&---				&0.002 711 198 384\\
50\phantom{.000}	&{0.00}2 006	&{0.000} 592\phantom{ 000}	&{0.000 642} 054\phantom{ 000} 	&{0.000 642} 496\phantom{ 000}		&-0.000 860	&---				&0.000 642 573 605\\
100\phantom{.00}	&{0.000} 708	&{0.000} 187\phantom{ 000}	&{0.000 220 6}05\phantom{ 000} 	&{0.000 220 6}83\phantom{ 000}		&-0.000 679	&---				&0.000 220 692 615\\
1000\phantom{.0}	&{0.000 0}22	&{0.000 006} 552		&{0.000 006 677} 055		&{0.000 006 677 3}02			&-0.000 112	&---		  	     	&0.000 006 677 311\\
\end{tabular}
\end{ruledtabular}
\footnotetext[1]{From the polynomial wavefunction in Section \ref{subsec:poly}}
\footnotetext[2]{Native result of Ref. \cite{SeidlPRA2007b}.}
\footnotetext[3]{Correlation energy form Ref. \cite{SeidlPRA2007b} relative to the UHF energy.}
\end{table*}
\end{turnpage}

\subsection{Second-order wavefunction}

To find the 4th- and 5th-order energies, we need the 2nd-order wavefunction.  This is given by
\begin{equation}
	\Phi^{(2)} (\theta) = \sum_{\ell=1}^\infty \frac{\left< \Phi^{(1)} \left| \hat{V} - E^{(1)} \right| \Phi_{\ell} \right>}{E_0 - E_{\ell}}
						\ \Phi_{\ell} (\theta)
\end{equation}
which yields
\begin{equation} \label{Phi-2}
\begin{split}
	\Phi^{(2)}(\theta) & = \frac{1}{4 \pi} \sum_{\ell_1,\ell_2=1}^\infty \sum_{\ell=|\ell_1-\ell_2|}^{\ell_1+\ell_2}	\\
						& \times \frac{2\ell_2+1}{\ell_1(\ell_1+1)\ell_2(\ell_2+1)} 
								\begin{pmatrix}
								\ell_1	&	\ell_2	&	\ell	\\
								0	&		0	&	0	\\
								\end{pmatrix}^2
							P_{\ell_2}(\cos\theta)
\end{split}
\end{equation}
Using the identity
\begin{equation} \label{Phi-2-sum}
	\sum_{\ell=1}^\infty \frac{2\ell+1}{\ell(\ell+1)} P_{\ell}(x) P_{\ell}(y) = - \ln \frac{(1-x)(1+y)}{4} - 1
\end{equation}
for $x \le y$, we eventually obtain
\begin{equation}
\begin{split}
	\Phi^{(2)}(u) 	& = (2\ml^2 -2\ml + 5) \phi_0 + (2\ml - 5) \phi_1(u) - \frac{\pi}{4}\\
			& + \frac{1}{\pi}\left[\Li_2 \left( \frac{1}{2} - \frac{u}{4R} \right) - 2 \Li_2 \left( -\frac{u}{2R} \right) \right]
\end{split}
\end{equation}
where $\Li_2$ is the dilogarithm function \cite{Lewin}.

\subsection{Fourth- and fifth-order energies}

The Wigner $2n+1$ rule and the closed-form expression of $\Phi^{(2)}$ yield the 4th- and 5th-order coefficients
\begin{align}
	\eps_4 	& = \frac{16}{3} (4\ml^3 - 42\ml^2 + 96\ml - 45) - 12\,\zeta(3)	\nonumber	\\
			& = -0.050\;275\;600\ldots													\\
	\eps_5	& = \frac{32}{3} (5\ml^4 - 90\ml^3 + 450\ml^2 - 660\ml + 252)	\nonumber	\\
			& + (216 - 80\ml) \zeta(3)						\nonumber	\\
			& = +0.013\;957\;832\ldots
\end{align}
where $\zeta$ is the Riemann zeta function.

\begin{center}
\begin{figure}
\caption{\label{fig:MPn}MP$n$ correlation energies as a function of $R$. The exact correlation energy is shown as the solid curve.}
	\includegraphics[width=0.48\textwidth]{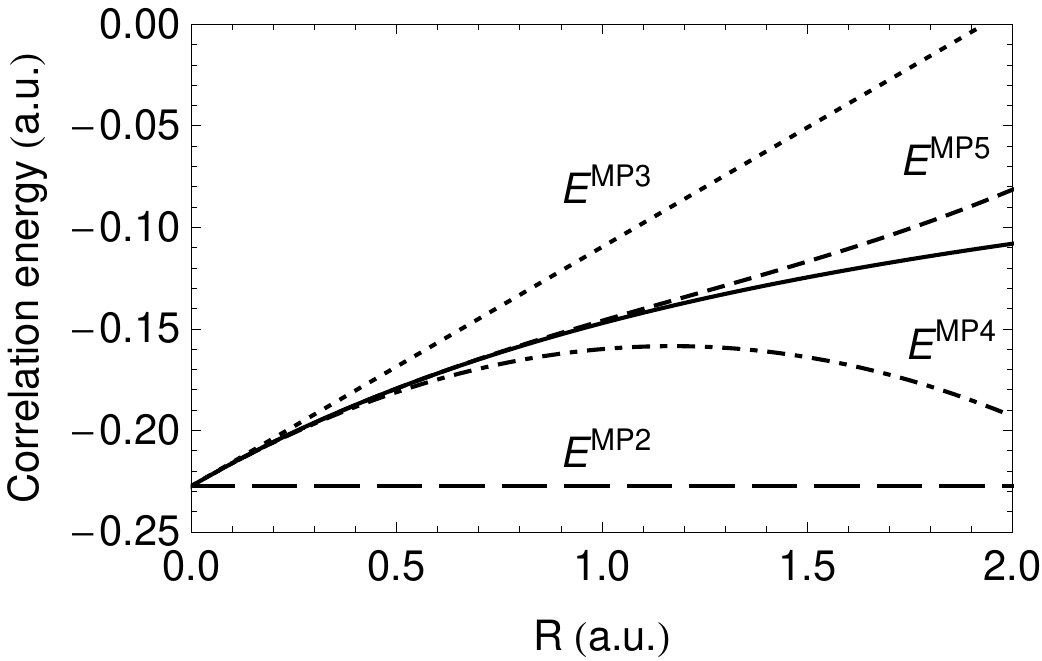}
\end{figure}
\end{center}

The MP$n$ correlation energies for various values of $R$ are reported in Table \ref{tab:Ecorr} and illustrated in Figure \ref{fig:MPn}.  The results show that MP4 and MP5 are very accurate for small $R$ and, indeed, the latter is reasonable up to $R\approx1$.

The MP expansion converges for radii $R$ within the radius of convergence
\begin{equation}
	R^{\rm cvg} = \lim_{n \to \infty} \left| \frac{\eps_n}{\eps_{n+1}} \right|
\end{equation}
From our results, it seems that $R^{\rm cvg} > 2$, but it is not possible to be more precise than this \cite{SeidlPRL2000, SeidlPRA2007b}.

\section{\label{sec:LargeR}Expansion for large R}

\subsection{Harmonic approximation}

For large $R$ (LR), the potential dominates the kinetic energy and the electrons tend to localize on opposite sides of the sphere.  The classical mechanical energy would be
\begin{equation}
	E^{e-e} = \frac{1}{2R}
\end{equation}
but, quantum mechanically, the kinetic energies of the electrons cannot vanish and each electron therefore maintains a zero-point oscillation around its equilibrium position with an angular frequency $\omega$.  Such phenomena are ubiquitous in strongly correlated systems, as demonstrated by Seidl and his co-workers \cite{SeidlPRA1999a, SeidlPRA1999b, SeidlPRL2000, SeidlPRA2007a, SeidlPRA2007b}.

In this limit, the supplementary angle $\xi = \pi - \theta$ is the natural coordinate and the Hamiltonian becomes
\begin{equation}\label{H-xi}
	\Hat{H} = - \frac{1}{R^2} \left( \frac{d^2}{d \xi^2} + \cot \xi \frac{d}{d \xi} \right) + \frac{1}{2 R} \sec \frac{\xi}{2}
\end{equation}
For small oscillations ($\xi \simeq 0$), the Taylor series
\begin{gather}
	\cot \xi = \xi^{-1} - \xi/3 - \xi^3/45 + \ldots		\\
	\sec (\xi/2) = 1 + \xi^2/8 + 5\xi^4/384 + 61\xi^6/46080 + \ldots
\end{gather}
yield the harmonic-oscillator Hamiltonian
\begin{equation} \label{eq:HarmHam}
	\Hat{H}^{\omega} = - \frac{1}{R^2} \left(\frac{d^2}{d \xi^2} + \frac{1}{\xi} \frac{d}{d\xi} \right) + \frac{1}{2R} \left(1+\frac{\xi^2}{8}\right)
\end{equation}
whose ground-state wavefunction and energy are
\begin{gather}
	\Phi_0^{\omega}(\xi) = \frac{1}{2 \sqrt{2} \pi R^{7/4}} \exp(-\sqrt{R}\,\xi^2/8)		\label{Psi-LargeR}	\\
	E^{\rm LR0} \equiv \mathcal{E}^{(0)} = \frac{1}{2 R} + \frac{1}{2R^{3/2}}	\label{E-Upsilon}
\end{gather}
The second term is the zero-point energy associated with harmonic oscillations of angular frequency $\omega = 1/R^{3/2}$ and it appears that this is the leading error in the UHF description at large $R$.

\subsection{First and second anharmonic corrections}

By analogy with the small-$R$ expansion \eqref{expansion-E}, we would like to construct a large-$R$ asymptotic expansion
\begin{align} \label{expansion-LargeR}
	E	& \sim \mathcal{E}^{(0)} + \mathcal{E}^{(1)} + \mathcal{E}^{(2)} + \ldots		\nonumber	\\
		& = \frac{\eta_1}{R} + \frac{\eta_2}{R^{3/2}} + \frac{\eta_3}{R^2} + \frac{\eta_4}{R^{5/2}} + \ldots
\end{align}
where we know $\eta_1 = \eta_2 = 1/2$.  The $n$th excited state of the Hamiltonian \eqref{eq:HarmHam} has the wavefunction and energy
\begin{gather}
	\Phi_n^{\omega}(\xi) = \Lag_n(\sqrt{R}\,\xi^2/4)\,\Phi_0^{\omega}(\xi)	\label{Psi-LargeR-ES}	\\
	E_n^{\omega} = \left( n + \frac{1}{2} \right) \omega						\label{Psi-LargeR-E}
\end{gather}
where $\Lag_n$ is the Laguerre polynomial of degree $n$ \cite{Abramowitz}.  The anharmonic corrections, $\mathcal{E}^{(1)}$ and $\mathcal{E}^{(2)}$, can be found \cite{BenderPR1969} using the perturbation operators
\begin{align}
	\Hat{W}^{(1)} & = - \frac{1}{R^2} \frac{\xi}{3} \frac{d}{d\xi} + \frac{1}{2 R} \frac{5\,\xi^4}{384}			\\
	\Hat{W}^{(2)} & = - \frac{1}{R^2} \frac{\xi^3}{45} \frac{d}{d\xi} + \frac{1}{2 R} \frac{61\,\xi^6}{46\,080}
\end{align}
The first-order correction is
\begin{align} \label{anharm1}
	\mathcal{E}^{(1)}	& = \left< \Phi_0^\omega \left| \hat{W}^{(1)} \right| \Phi_0^\omega \right>	\nonumber	\\
						& = 4\pi^2 R^4 \int_0^\infty \Phi_0^\omega(\xi)\,\hat{W}^{(1)} \Phi_0^\omega(\xi) \xi d\xi
\end{align}
and this yields $\eta_3 = - 1/8$ and therefore
\begin{equation}
	E^{\rm LR1} = \frac{1}{2 R} + \frac{1}{2R^{3/2}} - \frac{1}{8R^2}
\end{equation}
The second-order correction is
\begin{equation} \label{anharm2}
	\mathcal{E}^{(2)} = \sum_{n=1}^\infty \frac{\left<\Phi_0^\omega \left|\hat{W}^{(1)}\right| \Phi_n^\omega\right>^2}
						{E_0^\omega-E_n^\omega} + \left<\Phi_0^\omega \left|\hat{W}^{(2)}\right| \Phi_0^\omega\right>
\end{equation}
but because of the orthogonality and recurrence relations of Laguerre polynomials \cite{Abramowitz}, only the first two terms in the sum in \eqref{anharm2} are non-zero and one finds $\eta_4 = - 1/128$ and therefore
\begin{equation}
	E^{\rm LR2} = \frac{1}{2 R} + \frac{1}{2R^{3/2}} - \frac{1}{8R^2} - \frac{1}{128R^{5/2}}
\end{equation}

\begin{center}
\begin{figure}
\caption{\label{fig:LargeR}Correlation energies (relative to RHF) from $E^{\rm LR0}$ (dashed), $E^{\rm LR1}$ (dotted) and $E^{\rm LR2}$ (dot-dashed), $E^{\rm MP5}$ (small dash) 
and $E^{\rm exact}$ (solid) as a function of $R$.}
	\includegraphics[width=0.48\textwidth]{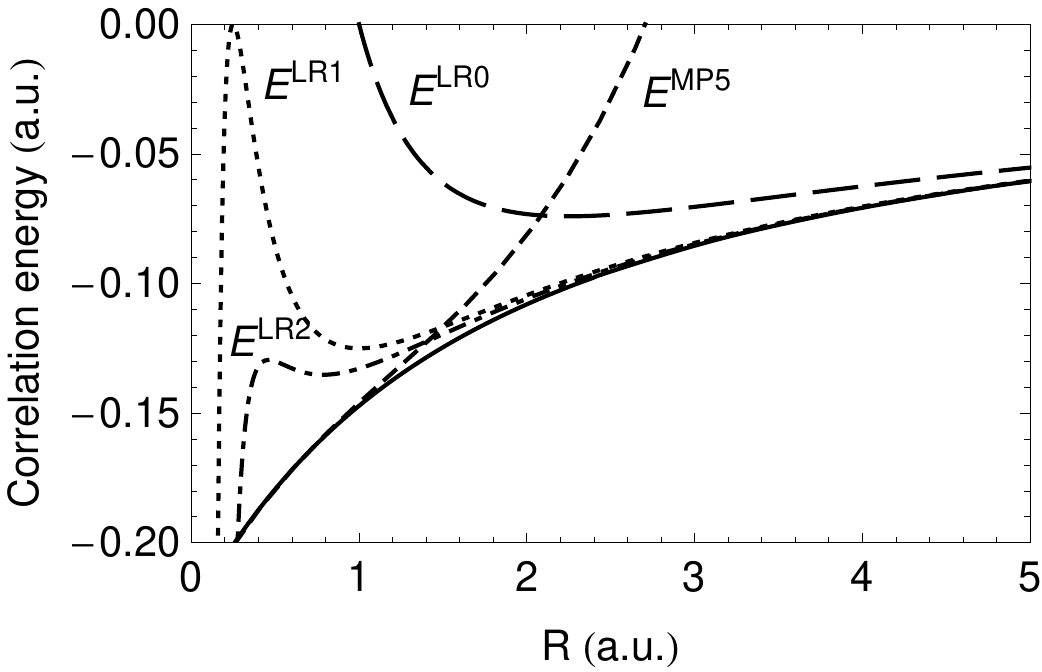}
\end{figure}
\end{center}

From the results in Table \ref{tab:Ecorr} and Figure \ref{fig:LargeR}, it seems that the asymptotic expansion converges toward the exact energy and is reasonably accurate for $R > 3$.

Through judicious use of the 5th-order truncation of \eqref{expansion-E} and the 2nd-order truncation of \eqref{expansion-LargeR}, one can predict satisfactory energies over a wide range of $R$ values.  However, there remains a region ($1 \lesssim R \lesssim 3$) where both the small-$R$ and large-$R$ solutions are inadequate.

\section{\label{sec:R12}Variational wavefunctions}

\subsection{\label{subsec:CI}Configuration interaction}

We begin with a configuration interaction (CI) treatment wherein the wavefunction is expanded as
\begin{equation}\label{Psi-CI}
	\Phi_L^{\rm CI} (\theta) = \sum_{\ell=0}^L T_\ell\,\Phi_\ell (\theta)
\end{equation}
in the Legendre polynomial basis set \eqref{L}.  The resulting energy $E_L^{\rm CI}$ is the lowest eigenvalue of the CI matrix
\begin{equation} \label{H-CI}
\begin{split}
	\left< \Phi_{\ell_1} \left| \hat{H} \right| \Phi_{\ell_2} \right> & = \frac{\ell_1(\ell_1+1)}{R^2} \delta_{\ell_1,\ell_2}		\\ 
			& + \frac{1}{R} \sum_{\ell=|\ell_1-\ell_2|}^{\ell_1+\ell_2} \sqrt{\frac{4 \pi}{2\ell+1}} \left< \ell_1\,\ell_2\,\ell \right>
\end{split}
\end{equation}
where $\left< \ell_1\,\ell_2\,\ell\right>$ is defined by \eqref{Y-W3J}.

The CI energy as the maximum angular momentum $L$ increases is reported in Table \ref{tab:CI}.  It converges very slowly and even $L=40$ yields an accuracy of only $10^{-4}$.  The reason for this slow convergence -- the failure of \eqref{Psi-CI} to satisfy the Kato cusp condition -- is well known.

\begin{table}
\caption{\label{tab:CI}Convergence of correlation energies with respect to the number $L$ of terms in the CI, R12-CI and Hylleraas wavefunctions. 
All results pertain to the sphere with $R=1$.}
\begin{ruledtabular}
\begin{tabular}{clll}
	$L$		&	\mc{1}{c}{CI}		&	\mc{1}{c}{R12-CI}		&	\mc{1}{c}{Polynomial}		\\
	\hline
	0		&	$-0.000\ 000$		&	$-0.147\ 180\ 860$		&	$-0.000\ 000\ 000\ 000$	\\
	1		&	$-0.131\ 665$		&	$-0.147\ 185\ 454$		&	$-0.147\ 180\ 859\ 845$	\\
	2		&	$-0.141\ 241$		&	$-0.147\ 202\ 916$		&	$-0.147\ 218\ 627\ 134$	\\
	3		&	$-0.144\ 065$		&	$-0.147\ 209\ 904$		&	$-0.147\ 218\ 930\ 072$	\\
	4		&	$-0.145\ 273$		&	$-0.147\ 213\ 200$		&	$-0.147\ 218\ 934\ 845$	\\
	5		&	$-0.145\ 900$		&	$-0.147\ 214\ 987$		&	$-0.147\ 218\ 935\ 941$	\\
	10		&	$-0.146\ 847$		&	$-0.147\ 217\ 796$		&	$-0.147\ 218\ 935\ 944$	\\
	15		&	$-0.147\ 047$		&	$-0.147\ 218\ 405$		&	$-0.147\ 218\ 935\ 944$	\\
	20		&	$-0.147\ 120$		&	$-0.147\ 218\ 631$		&	$-0.147\ 218\ 935\ 944$	\\
	25		&	$-0.147\ 155$		&	$-0.147\ 218\ 738$		&	$-0.147\ 218\ 935\ 944$	\\
	30		&	$-0.147\ 174$		&	$-0.147\ 218\ 797$		&	$-0.147\ 218\ 935\ 944$	\\
	35		&	$-0.147\ 186$		&	$-0.147\ 218\ 833$		&	$-0.147\ 218\ 935\ 944$	\\
	40		&	$-0.147\ 194$		&	$-0.147\ 218\ 857$		&	$-0.147\ 218\ 935\ 944$	\\
\end{tabular}
\end{ruledtabular}
\end{table}

\subsection{\label{subsec:Ref-R12} Hylleraas}

The simplest possible wavefunction with a cusp is
\begin{equation}\label{PhiHy}
	\Phi^{\rm Hy} = 1 + \gamma\,u
\end{equation}
which has an explicit linear dependence on the interelectronic distance $u$. 
Kato proved \cite{Kato1957} that $\gamma = 1/2$ in normal singlet states but, because our electrons are confined to a sphere, this $\gamma$ does not apply (see below).

Using the partial-wave expansion
\begin{equation}\label{r12-expansion}
	u = R \sum_{\ell=0}^\infty \left( \frac{1}{2\ell+3} - \frac{1}{2\ell-1} \right) P_{\ell} (\cos \theta)
\end{equation}
one finds that the energy is
\begin{equation}\label{E-R12}
	E^{\rm Hy}(\gamma) = \frac{6 + 3\gamma(\gamma + 4)R + 8\gamma^2 R^2}
								{2R(3 + 8\gamma R + 6\gamma^2 R^2)}
\end{equation}
and minimizing this with respect to $\gamma$ yields
\begin{gather}
	\gamma^{\rm opt} = \frac{12}{9 - 12 R + \sqrt{81 + 72R + 48R^2}}	\label{gamma-opt}\\
	E^{\rm Hy} = \frac{9 + 12 R - \sqrt{81 + 72R + 48R^2}}{8 R^2}		\label{E-R12-opt}
\end{gather}
Correlation energies obtained from \eqref{E-R12-opt} for several values of $R$ are reported in Table \ref{tab:Ecorr}.  Despite the simplicity of the wavefunction, its energies are surprisingly good with a maximum deviation of $0.003$ for large $R$ and $0.005$ for small $R$.  As $R$ tends to zero, the correlation energy approaches $-0.222\,222$, which is close to the exact value $-0.227\,411$.  However, as $R$ becomes large, one can show that $E^{\rm Hy} \sim 1/(1.58 R)$ which lies between the RHF and UHF energies.  The Hylleraas wavefunction is thus a useful alternative to the small- and large-$R$ solutions in the problematic intermediate region ($1 \lesssim R \lesssim 3$) with errors of 0.000, 0.0011 and 0.0021 for $R =$ 1, 2 and 3, respectively.

\subsection{R12-CI}

Using the Hylleraas wavefunction \eqref{PhiHy} as the reference for a CI expansion yields the R12-CI wavefunction
\begin{equation}\label{Psi-R12CI}
	\Phi_L^{\text{R12-CI}} (\theta) = \Hat{P}\,\Phi^{\rm Hy}(\theta) + \sum_{\ell=1}^L T_\ell \,\Phi_\ell(\theta)
\end{equation}
where
\begin{equation}\label{projector}
	\hat{P} = \hat{I} - \sum_{\ell=1}^\infty \left| \Phi_\ell \right> \left< \Phi_\ell \right| 
\end{equation}
is a projection operator that ensures orthogonality between the reference wavefunction and the excited determinants and $\Hat{I}$ is the identity operator.  The coupling coefficients between two basis functions are the same as those for the conventional CI calculation \eqref{H-CI} but with a correction for the matrix element
\begin{equation}
\begin{split}
	\left< \Phi^{\rm Hy} \left| \Hat{H} \right| \Phi_\ell \right> & = \left< \Phi^{\rm RHF} \left| \Hat{H} \right| \Phi_\ell \right>	\\
					&+ \gamma \frac{\ell(\ell+1)}{R\sqrt{2\ell+1}} \left( \frac{1}{2\ell+3} - \frac{1}{2\ell-1} \right)
\end{split}
\end{equation}
involving the ground state and the excited determinants.  It is no longer possible to optimize $\gamma$ in closed form so we used the value given by \eqref{gamma-opt}.

As Table \ref{tab:CI} shows, the R12-CI energies converge much more rapidly with $L$ than the CI energies and, for example, $E_2^{\text{R12-CI}}$ is more accurate than $E_{40}^{\rm CI}$.  This illustrates the importance of including a term that is linear in $u$.  However, although this term enhances the \emph{initial} convergence rate, the asymptotic behavior of the CI and R12-CI schemes are identical.  Therefore, we now investigate the effect of including higher-order $u$ terms.

\subsection{\label{subsec:poly}Polynomial}

In terms of the distance $u$, the Hamiltonian is
\begin{equation}\label{H-r12}
	\Hat{H} = \left( \frac{u^2}{4R^2} - 1 \right) \frac{d^2}{du^2} + \left( \frac{3u}{4R^2} - \frac{1}{u} \right) \frac{d}{du} + \frac{1}{u}
\end{equation}
and a Kato-like analysis \cite{Kato1957} reveals the cusp condition
\begin{equation}\label{H-cusp}
	\frac{\Phi'(0)}{\Phi(0)} = 1
\end{equation}
which deviates from the normal value of 1/2 \cite{SeidlPRA2007b}.

The natural generalization of the Hylleraas wavefunction \eqref{PhiHy} is a polynomial and it is convenient to select the orthonormal basis of Jacobi polynomials \cite{Abramowitz}
\begin{equation}
	\Xi_\ell (u) = \frac{\sqrt{\ell+1}}{4 \pi R^2} P_\ell^{(1,0)} \left(1-\frac{u}{R} \right)
\end{equation}
and write the wavefunction as
\begin{equation}
	\Phi_L^{\rm poly} = \sum_{\ell=0}^L c_\ell \,\Xi_\ell(u)
\end{equation}
The energy $E_L^{\rm poly}$ is the lowest eigenvalue of the matrix
\begin{equation}
	\left< \Xi_i \left| \hat{H} \right| \Xi_j \right> = \frac{(m^2 - 1)(\alpha m + \delta_{i,j})}{4R^2} + \frac{\alpha m}{R}
\end{equation}
where $m=\min(i,j)$ and $\alpha = \sqrt{\frac{\min(i,j)}{\max(i,j)}}$.

Table \ref{tab:CI} reveals the remarkable convergence of $E_L^{\rm poly}$.  Using $L=40$ and $R=1$, for example, we find
\begin{multline}
	E_{40}^{\rm poly} = 0.852\;781\;065\;056\;462\;665\;400\;437\;966\;038\;710	\\
			264\;283\;589\;518\;406\;360\;162\;484\;313\;983
\end{multline}
The convergence is slower for larger values of $R$, but still impressive.  For example, using $L=40$ and $R = 1000$, the energy is still correct to 49 digits.  The ease with which we can obtain these Schr\"odinger eigenvalues can be traced to the fact that the polynomial basis efficiently models all of the singularities (the 1st-order cusp, the third-order cusp, etc.) in the exact wavefunction.

In recent work \cite{SeidlPRA2007b}, Seidl reported correlation energies based on his numerical integration of the Schr{\"o}dinger equation from $\theta = 0$ to $\pi$ using \eqref{H-theta} and we have included these in Table \ref{tab:Ecorr}. It appears that some of his energies for small $R$ are slightly inaccurate.

\section{Conclusion}

In this article, we have reported results for the ground state of a simple two-electron system that is described by a single parameter $R$.  Although we cannot solve its Schr\"odinger equation in closed form, we have found accurate wavefunctions and energies for small $R$ (the weakly correlated limit) and large $R$ (the strongly correlated limit).  For $R \ll 1$, M{\o}ller-Plesset perturbation theory yields results close to the exact solution; for $R \gg 1$, accurate results can be found by considering the zero-point oscillations of the appropriate Wigner molecule.

We have also explored variational schemes that yield satisfactory results for all $R$.  In particular, we have discovered a polynomial wavefunction that easily yields results of any required accuracy.

We believe that our results will be useful in the future development of accurate correlation functionals within density-functional theory \cite{SunJCTC2009, GoriGiorgiJCTC2009} and intracule functional theory \cite{GillPCCP2005, DumontPCCP2007, CrittendenJCP2007a, CrittendenJCP2007b, BernardPCCP2008, PearsonJCP2009}.

\begin{acknowledgments}
PMWG thanks the APAC Merit Allocation Scheme for a grant of supercomputer time and Australian Research Council (Grant DP0664466) for funding. 
We also thank Yves Bernard for helpful comments on the manuscript, and fruitful discussions.
\end{acknowledgments}


\begin{thebibliography}{49}
\expandafter\ifx\csname natexlab\endcsname\relax\def\natexlab#1{#1}\fi
\expandafter\ifx\csname bibnamefont\endcsname\relax
  \def\bibnamefont#1{#1}\fi
\expandafter\ifx\csname bibfnamefont\endcsname\relax
  \def\bibfnamefont#1{#1}\fi
\expandafter\ifx\csname citenamefont\endcsname\relax
  \def\citenamefont#1{#1}\fi
\expandafter\ifx\csname url\endcsname\relax
  \def\url#1{\texttt{#1}}\fi
\expandafter\ifx\csname urlprefix\endcsname\relax\def\urlprefix{URL }\fi
\providecommand{\bibinfo}[2]{#2}
\providecommand{\eprint}[2][]{\url{#2}}

\bibitem[{\citenamefont{Hohenberg and Kohn}(1964)}]{HohenbergPRB1964}
\bibinfo{author}{\bibfnamefont{P.}~\bibnamefont{Hohenberg}} \bibnamefont{and}
  \bibinfo{author}{\bibfnamefont{W.}~\bibnamefont{Kohn}},
  \bibinfo{journal}{Phys. Rev. B} \textbf{\bibinfo{volume}{136}},
  \bibinfo{pages}{864} (\bibinfo{year}{1964}).

\bibitem[{\citenamefont{Kohn and Sham}(1965)}]{KohnPRA1965}
\bibinfo{author}{\bibfnamefont{W.}~\bibnamefont{Kohn}} \bibnamefont{and}
  \bibinfo{author}{\bibfnamefont{L.}~\bibnamefont{Sham}},
  \bibinfo{journal}{Phys. Rev. A} \textbf{\bibinfo{volume}{140}},
  \bibinfo{pages}{1133} (\bibinfo{year}{1965}).

\bibitem[{\citenamefont{Parr and Yang}(1989)}]{ParrYang}
\bibinfo{author}{\bibfnamefont{R.~G.} \bibnamefont{Parr}} \bibnamefont{and}
  \bibinfo{author}{\bibfnamefont{W.}~\bibnamefont{Yang}},
  \emph{\bibinfo{title}{Density Functional Theory for Atoms and Molecules}}
  (\bibinfo{publisher}{Oxford University Press}, \bibinfo{year}{1989}).

\bibitem[{\citenamefont{Kestner and Sinanoglu}(1962)}]{KestnerPhysRev1962}
\bibinfo{author}{\bibfnamefont{N.~R.} \bibnamefont{Kestner}} \bibnamefont{and}
  \bibinfo{author}{\bibfnamefont{O.}~\bibnamefont{Sinanoglu}},
  \bibinfo{journal}{Phys. Rev.} \textbf{\bibinfo{volume}{128}},
  \bibinfo{pages}{2687} (\bibinfo{year}{1962}).

\bibitem[{\citenamefont{Kais et~al.}(1989)\citenamefont{Kais, Herschbach, and
  Levine}}]{KaisJCP1989}
\bibinfo{author}{\bibfnamefont{S.}~\bibnamefont{Kais}},
  \bibinfo{author}{\bibfnamefont{D.~R.} \bibnamefont{Herschbach}},
  \bibnamefont{and} \bibinfo{author}{\bibfnamefont{R.~D.}
  \bibnamefont{Levine}}, \bibinfo{journal}{J. Chem. Phys}
  \textbf{\bibinfo{volume}{91}}, \bibinfo{pages}{7791} (\bibinfo{year}{1989}).

\bibitem[{\citenamefont{Taut}(1993)}]{TautPRA1993}
\bibinfo{author}{\bibfnamefont{M.}~\bibnamefont{Taut}}, \bibinfo{journal}{Phys.
  Rev. A} \textbf{\bibinfo{volume}{48}}, \bibinfo{pages}{3561}
  (\bibinfo{year}{1993}).

\bibitem[{\citenamefont{Alavi}(2000)}]{AlaviJCP2000}
\bibinfo{author}{\bibfnamefont{A.}~\bibnamefont{Alavi}}, \bibinfo{journal}{J.
  Chem. Phys.} \textbf{\bibinfo{volume}{113}}, \bibinfo{pages}{7735}
  (\bibinfo{year}{2000}).

\bibitem[{\citenamefont{Thompson and Alavi}(2002)}]{ThompsonPRB2002}
\bibinfo{author}{\bibfnamefont{D.~C.} \bibnamefont{Thompson}} \bibnamefont{and}
  \bibinfo{author}{\bibfnamefont{A.}~\bibnamefont{Alavi}},
  \bibinfo{journal}{Phys. Rev. B} \textbf{\bibinfo{volume}{66}},
  \bibinfo{pages}{235118} (\bibinfo{year}{2002}).

\bibitem[{\citenamefont{Thompson and Alavi}(2005)}]{ThompsonJCP2005}
\bibinfo{author}{\bibfnamefont{D.~C.} \bibnamefont{Thompson}} \bibnamefont{and}
  \bibinfo{author}{\bibfnamefont{A.}~\bibnamefont{Alavi}}, \bibinfo{journal}{J.
  Chem. Phys.} \textbf{\bibinfo{volume}{122}}, \bibinfo{pages}{124107}
  (\bibinfo{year}{2005}).

\bibitem[{\citenamefont{Thompson and Alavi}(2004)}]{ThompsonPRB2004}
\bibinfo{author}{\bibfnamefont{D.~C.} \bibnamefont{Thompson}} \bibnamefont{and}
  \bibinfo{author}{\bibfnamefont{A.}~\bibnamefont{Alavi}},
  \bibinfo{journal}{Phys. Rev. B} \textbf{\bibinfo{volume}{69}},
  \bibinfo{pages}{201302} (\bibinfo{year}{2004}).

\bibitem[{\citenamefont{Wigner}(1934)}]{WignerPR1934}
\bibinfo{author}{\bibfnamefont{E.}~\bibnamefont{Wigner}},
  \bibinfo{journal}{Phys. Rev.} \textbf{\bibinfo{volume}{46}},
  \bibinfo{pages}{1002} (\bibinfo{year}{1934}).

\bibitem[{\citenamefont{Ezra and Berry}(1982)}]{EzraPRA1982}
\bibinfo{author}{\bibfnamefont{G.~S.} \bibnamefont{Ezra}} \bibnamefont{and}
  \bibinfo{author}{\bibfnamefont{R.~S.} \bibnamefont{Berry}},
  \bibinfo{journal}{Phys. Rev. A} \textbf{\bibinfo{volume}{25}},
  \bibinfo{pages}{1513} (\bibinfo{year}{1982}).

\bibitem[{\citenamefont{Ezra and Berry}(1983)}]{EzraPRA1983}
\bibinfo{author}{\bibfnamefont{G.~S.} \bibnamefont{Ezra}} \bibnamefont{and}
  \bibinfo{author}{\bibfnamefont{R.~S.} \bibnamefont{Berry}},
  \bibinfo{journal}{Phys. Rev. A} \textbf{\bibinfo{volume}{28}},
  \bibinfo{pages}{1989} (\bibinfo{year}{1983}).

\bibitem[{\citenamefont{Ojha and Berry}(1987)}]{OjhaPRA1987}
\bibinfo{author}{\bibfnamefont{P.~C.} \bibnamefont{Ojha}} \bibnamefont{and}
  \bibinfo{author}{\bibfnamefont{R.~S.} \bibnamefont{Berry}},
  \bibinfo{journal}{Phys. Rev. A} \textbf{\bibinfo{volume}{36}},
  \bibinfo{pages}{1575} (\bibinfo{year}{1987}).

\bibitem[{\citenamefont{Hinde and Berry}(1990)}]{HindePRA1990}
\bibinfo{author}{\bibfnamefont{R.~J.} \bibnamefont{Hinde}} \bibnamefont{and}
  \bibinfo{author}{\bibfnamefont{R.~S.} \bibnamefont{Berry}},
  \bibinfo{journal}{Phys. Rev. A} \textbf{\bibinfo{volume}{42}},
  \bibinfo{pages}{2259} (\bibinfo{year}{1990}).

\bibitem[{\citenamefont{Warner and Berry}(1985)}]{WarnerNature1985}
\bibinfo{author}{\bibfnamefont{J.~W.} \bibnamefont{Warner}} \bibnamefont{and}
  \bibinfo{author}{\bibfnamefont{R.~S.} \bibnamefont{Berry}},
  \bibinfo{journal}{Nature} \textbf{\bibinfo{volume}{313}},
  \bibinfo{pages}{160} (\bibinfo{year}{1985}).

\bibitem[{\citenamefont{Seidl}(2007{\natexlab{a}})}]{SeidlPRA2007b}
\bibinfo{author}{\bibfnamefont{M.}~\bibnamefont{Seidl}},
  \bibinfo{journal}{Phys. Rev. A} \textbf{\bibinfo{volume}{75}},
  \bibinfo{pages}{062506} (\bibinfo{year}{2007}{\natexlab{a}}).

\bibitem[{\citenamefont{Seidl et~al.}(2000)\citenamefont{Seidl, Perdew, and
  Kurth}}]{SeidlPRL2000}
\bibinfo{author}{\bibfnamefont{M.}~\bibnamefont{Seidl}},
  \bibinfo{author}{\bibfnamefont{J.~P.} \bibnamefont{Perdew}},
  \bibnamefont{and} \bibinfo{author}{\bibfnamefont{S.}~\bibnamefont{Kurth}},
  \bibinfo{journal}{Phys. Rev. Lett.} \textbf{\bibinfo{volume}{84}},
  \bibinfo{pages}{5070} (\bibinfo{year}{2000}).

\bibitem[{\citenamefont{M{\o}ller and Plesset}(1934)}]{MollerPhysRev1934}
\bibinfo{author}{\bibfnamefont{C.}~\bibnamefont{M{\o}ller}} \bibnamefont{and}
  \bibinfo{author}{\bibfnamefont{M.~S.} \bibnamefont{Plesset}},
  \bibinfo{journal}{Phys. Rev.} \textbf{\bibinfo{volume}{46}},
  \bibinfo{pages}{618} (\bibinfo{year}{1934}).

\bibitem[{\citenamefont{Hylleraas}(1964)}]{HylleraasAdvQuantumChem1964}
\bibinfo{author}{\bibfnamefont{E.~A.} \bibnamefont{Hylleraas}},
  \bibinfo{journal}{Adv. Quantum Chem.} \textbf{\bibinfo{volume}{1}},
  \bibinfo{pages}{1} (\bibinfo{year}{1964}).

\bibitem[{\citenamefont{Kutzelnigg}(1985)}]{KutzelniggTheorChemAcc1985}
\bibinfo{author}{\bibfnamefont{W.}~\bibnamefont{Kutzelnigg}},
  \bibinfo{journal}{Theor. Chim. Acta} \textbf{\bibinfo{volume}{68}},
  \bibinfo{pages}{445} (\bibinfo{year}{1985}).

\bibitem[{\citenamefont{Klopper and Kutzelnigg}(1987)}]{KlopperCPL1987}
\bibinfo{author}{\bibfnamefont{W.}~\bibnamefont{Klopper}} \bibnamefont{and}
  \bibinfo{author}{\bibfnamefont{W.}~\bibnamefont{Kutzelnigg}},
  \bibinfo{journal}{Chem. Phys. Lett.} \textbf{\bibinfo{volume}{134}},
  \bibinfo{pages}{17} (\bibinfo{year}{1987}).

\bibitem[{\citenamefont{Klopper and Kutzelnigg}(1990)}]{KlopperJPhysChem1990}
\bibinfo{author}{\bibfnamefont{W.}~\bibnamefont{Klopper}} \bibnamefont{and}
  \bibinfo{author}{\bibfnamefont{W.}~\bibnamefont{Kutzelnigg}},
  \bibinfo{journal}{J. Phys. Chem.} \textbf{\bibinfo{volume}{94}},
  \bibinfo{pages}{5625} (\bibinfo{year}{1990}).

\bibitem[{\citenamefont{Kutzelnigg and
  Klopper}(1991)}]{KutzelniggJChemPhys1991}
\bibinfo{author}{\bibfnamefont{W.}~\bibnamefont{Kutzelnigg}} \bibnamefont{and}
  \bibinfo{author}{\bibfnamefont{W.}~\bibnamefont{Klopper}},
  \bibinfo{journal}{J. Chem. Phys.} \textbf{\bibinfo{volume}{94}},
  \bibinfo{pages}{1985} (\bibinfo{year}{1991}).

\bibitem[{\citenamefont{Kato}(1957)}]{Kato1957}
\bibinfo{author}{\bibfnamefont{T.}~\bibnamefont{Kato}},
  \bibinfo{journal}{Commun. Pure Appl. Math.} \textbf{\bibinfo{volume}{10}},
  \bibinfo{pages}{151} (\bibinfo{year}{1957}).

\bibitem[{\citenamefont{Pack and Byers~Brown}(1966)}]{PackJChemPhys1966}
\bibinfo{author}{\bibfnamefont{R.~T.} \bibnamefont{Pack}} \bibnamefont{and}
  \bibinfo{author}{\bibfnamefont{W.}~\bibnamefont{Byers~Brown}},
  \bibinfo{journal}{J. Chem. Phys.} \textbf{\bibinfo{volume}{45}},
  \bibinfo{pages}{556} (\bibinfo{year}{1966}).

\bibitem[{\citenamefont{Szabo and Ostlund}(1989)}]{Szabo}
\bibinfo{author}{\bibfnamefont{A.}~\bibnamefont{Szabo}} \bibnamefont{and}
  \bibinfo{author}{\bibfnamefont{N.~S.} \bibnamefont{Ostlund}},
  \emph{\bibinfo{title}{Modern Quantum Chemistry : Introduction to Advanced
  Structure Theory}} (\bibinfo{publisher}{Dover publications Inc., Mineola,
  New-York}, \bibinfo{year}{1989}).

\bibitem[{\citenamefont{Arfken}(1966)}]{Arfken}
\bibinfo{author}{\bibfnamefont{G.~B.} \bibnamefont{Arfken}},
  \emph{\bibinfo{title}{Mathematical Methods for Physicists}}
  (\bibinfo{publisher}{Academic Press, New-York}, \bibinfo{year}{1966}).

\bibitem[{\citenamefont{Abramowitz and Stegun}(1972)}]{Abramowitz}
\bibinfo{author}{\bibfnamefont{M.}~\bibnamefont{Abramowitz}} \bibnamefont{and}
  \bibinfo{author}{\bibfnamefont{I.~A.} \bibnamefont{Stegun}},
  \emph{\bibinfo{title}{Handbook of Mathematical Functions with Formulas,
  Graphs and Mathematical Tables}} (\bibinfo{publisher}{Dover publications
  Inc., New-York}, \bibinfo{year}{1972}).

\bibitem[{\citenamefont{Cizek and Paldus}(1967)}]{CizekJCP1967}
\bibinfo{author}{\bibfnamefont{J.}~\bibnamefont{Cizek}} \bibnamefont{and}
  \bibinfo{author}{\bibfnamefont{L.}~\bibnamefont{Paldus}},
  \bibinfo{journal}{J. Chem. Phys.} \textbf{\bibinfo{volume}{47}},
  \bibinfo{pages}{3976} (\bibinfo{year}{1967}).

\bibitem[{\citenamefont{Paldus and Cizek}(1970)}]{CizekJCP1970}
\bibinfo{author}{\bibfnamefont{L.}~\bibnamefont{Paldus}} \bibnamefont{and}
  \bibinfo{author}{\bibfnamefont{J.}~\bibnamefont{Cizek}}, \bibinfo{journal}{J.
  Chem. Phys.} \textbf{\bibinfo{volume}{52}}, \bibinfo{pages}{2919}
  (\bibinfo{year}{1970}).

\bibitem[{\citenamefont{Seeger and Pople}(1977)}]{SeegerJCP1977}
\bibinfo{author}{\bibfnamefont{R.}~\bibnamefont{Seeger}} \bibnamefont{and}
  \bibinfo{author}{\bibfnamefont{J.}~\bibnamefont{Pople}}, \bibinfo{journal}{J.
  Chem. Phys.} \textbf{\bibinfo{volume}{66}}, \bibinfo{pages}{3045}
  (\bibinfo{year}{1977}).

\bibitem[{\citenamefont{Edmonds}(1957)}]{Edmonds}
\bibinfo{author}{\bibfnamefont{A.~R.} \bibnamefont{Edmonds}},
  \emph{\bibinfo{title}{Angular Momentum in Quantum Mechanics}}
  (\bibinfo{publisher}{Princeton University Press}, \bibinfo{year}{1957}).

\bibitem[{\citenamefont{Slater}(1960)}]{Slater}
\bibinfo{author}{\bibfnamefont{J.~C.} \bibnamefont{Slater}},
  \emph{\bibinfo{title}{Quantum Theory of Atomic Structures}},
  vol.~\bibinfo{volume}{2} of \emph{\bibinfo{series}{International Series in
  Pure and Applied Physics}} (\bibinfo{publisher}{McGraw-Hill Book Compagny,
  Inc.}, \bibinfo{year}{1960}).

\bibitem[{\citenamefont{Helgaker et~al.}(2000)\citenamefont{Helgaker,
  J{\o}rgensen, and Olsen}}]{Helgaker}
\bibinfo{author}{\bibfnamefont{T.}~\bibnamefont{Helgaker}},
  \bibinfo{author}{\bibfnamefont{P.}~\bibnamefont{J{\o}rgensen}},
  \bibnamefont{and} \bibinfo{author}{\bibfnamefont{J.}~\bibnamefont{Olsen}},
  \emph{\bibinfo{title}{Molecular Electronic-Structure Theory}}
  (\bibinfo{publisher}{John Wiley \& Sons, Ltd.}, \bibinfo{year}{2000}).

\bibitem[{\citenamefont{Gill and O'Neill}(2005)}]{GillJCP2005}
\bibinfo{author}{\bibfnamefont{P.~M.~W.} \bibnamefont{Gill}} \bibnamefont{and}
  \bibinfo{author}{\bibfnamefont{D.~P.} \bibnamefont{O'Neill}},
  \bibinfo{journal}{J. Chem. Phys.} \textbf{\bibinfo{volume}{122}},
  \bibinfo{pages}{094110} (\bibinfo{year}{2005}).

\bibitem[{\citenamefont{Lewin}(1958)}]{Lewin}
\bibinfo{author}{\bibfnamefont{L.}~\bibnamefont{Lewin}},
  \emph{\bibinfo{title}{Dilogarithms and Associated Functions}}
  (\bibinfo{publisher}{London: Macdonald}, \bibinfo{year}{1958}).

\bibitem[{\citenamefont{Seidl et~al.}(1999)\citenamefont{Seidl, Perdew, and
  Levy}}]{SeidlPRA1999a}
\bibinfo{author}{\bibfnamefont{M.}~\bibnamefont{Seidl}},
  \bibinfo{author}{\bibfnamefont{J.~P.} \bibnamefont{Perdew}},
  \bibnamefont{and} \bibinfo{author}{\bibfnamefont{M.}~\bibnamefont{Levy}},
  \bibinfo{journal}{Phys. Rev. A} \textbf{\bibinfo{volume}{59}},
  \bibinfo{pages}{51} (\bibinfo{year}{1999}).

\bibitem[{\citenamefont{Seidl}(1999)}]{SeidlPRA1999b}
\bibinfo{author}{\bibfnamefont{M.}~\bibnamefont{Seidl}},
  \bibinfo{journal}{Phys. Rev. A} \textbf{\bibinfo{volume}{60}},
  \bibinfo{pages}{4387} (\bibinfo{year}{1999}).

\bibitem[{\citenamefont{Seidl}(2007{\natexlab{b}})}]{SeidlPRA2007a}
\bibinfo{author}{\bibfnamefont{M.}~\bibnamefont{Seidl}},
  \bibinfo{journal}{Phys. Rev. A} \textbf{\bibinfo{volume}{75}},
  \bibinfo{pages}{042511} (\bibinfo{year}{2007}{\natexlab{b}}).

\bibitem[{\citenamefont{Bender and Wu}(1969)}]{BenderPR1969}
\bibinfo{author}{\bibfnamefont{C.~M.} \bibnamefont{Bender}} \bibnamefont{and}
  \bibinfo{author}{\bibfnamefont{T.~T.} \bibnamefont{Wu}},
  \bibinfo{journal}{Phys. Rev.} \textbf{\bibinfo{volume}{184}},
  \bibinfo{pages}{1231} (\bibinfo{year}{1969}).

\bibitem[{\citenamefont{Sun}(2009)}]{SunJCTC2009}
\bibinfo{author}{\bibfnamefont{J.}~\bibnamefont{Sun}}, \bibinfo{journal}{J.
  Chem. Theor. Comput.} \textbf{\bibinfo{volume}{5}}, \bibinfo{pages}{708}
  (\bibinfo{year}{2009}).

\bibitem[{\citenamefont{Gori-Giorgi et~al.}(2009)\citenamefont{Gori-Giorgi,
  Vignale, and Seidl}}]{GoriGiorgiJCTC2009}
\bibinfo{author}{\bibfnamefont{P.}~\bibnamefont{Gori-Giorgi}},
  \bibinfo{author}{\bibfnamefont{G.}~\bibnamefont{Vignale}}, \bibnamefont{and}
  \bibinfo{author}{\bibfnamefont{M.}~\bibnamefont{Seidl}}, \bibinfo{journal}{J.
  Chem. Theor. Comput.} \textbf{\bibinfo{volume}{5}}, \bibinfo{pages}{743}
  (\bibinfo{year}{2009}).

\bibitem[{\citenamefont{Gill et~al.}(2005)\citenamefont{Gill, Crittenden,
  O'Neill, and Besley}}]{GillPCCP2005}
\bibinfo{author}{\bibfnamefont{P.~M.~W.} \bibnamefont{Gill}},
  \bibinfo{author}{\bibfnamefont{D.~L.} \bibnamefont{Crittenden}},
  \bibinfo{author}{\bibfnamefont{D.~P.} \bibnamefont{O'Neill}},
  \bibnamefont{and} \bibinfo{author}{\bibfnamefont{N.~A.}
  \bibnamefont{Besley}}, \bibinfo{journal}{Phys. Chem. Chem. Phys.}
  \textbf{\bibinfo{volume}{8}}, \bibinfo{pages}{15} (\bibinfo{year}{2005}).

\bibitem[{\citenamefont{Dumont et~al.}(2007)\citenamefont{Dumont, Crittenden,
  and Gill}}]{DumontPCCP2007}
\bibinfo{author}{\bibfnamefont{E.~E.} \bibnamefont{Dumont}},
  \bibinfo{author}{\bibfnamefont{D.~L.} \bibnamefont{Crittenden}},
  \bibnamefont{and} \bibinfo{author}{\bibfnamefont{P.~M.~W.}
  \bibnamefont{Gill}}, \bibinfo{journal}{Phys. Chem. Chem. Phys.}
  \textbf{\bibinfo{volume}{9}}, \bibinfo{pages}{5340} (\bibinfo{year}{2007}).

\bibitem[{\citenamefont{Crittenden and Gill}(2007)}]{CrittendenJCP2007a}
\bibinfo{author}{\bibfnamefont{D.~L.} \bibnamefont{Crittenden}}
  \bibnamefont{and} \bibinfo{author}{\bibfnamefont{P.~M.~W.}
  \bibnamefont{Gill}}, \bibinfo{journal}{J. Chem. Phys.}
  \textbf{\bibinfo{volume}{127}}, \bibinfo{pages}{014101}
  (\bibinfo{year}{2007}).

\bibitem[{\citenamefont{Crittenden et~al.}(2007)\citenamefont{Crittenden,
  Dumont, and Gill}}]{CrittendenJCP2007b}
\bibinfo{author}{\bibfnamefont{D.~L.} \bibnamefont{Crittenden}},
  \bibinfo{author}{\bibfnamefont{E.~E.} \bibnamefont{Dumont}},
  \bibnamefont{and} \bibinfo{author}{\bibfnamefont{P.~M.~W.}
  \bibnamefont{Gill}}, \bibinfo{journal}{J. Chem. Phys.}
  \textbf{\bibinfo{volume}{127}}, \bibinfo{pages}{141103}
  (\bibinfo{year}{2007}).

\bibitem[{\citenamefont{Bernard et~al.}(2008)\citenamefont{Bernard, Crittenden,
  and Gill}}]{BernardPCCP2008}
\bibinfo{author}{\bibfnamefont{Y.~A.} \bibnamefont{Bernard}},
  \bibinfo{author}{\bibfnamefont{D.~L.} \bibnamefont{Crittenden}},
  \bibnamefont{and} \bibinfo{author}{\bibfnamefont{P.~M.~W.}
  \bibnamefont{Gill}}, \bibinfo{journal}{Phys. Chem. Chem. Phys.}
  \textbf{\bibinfo{volume}{10}}, \bibinfo{pages}{3447} (\bibinfo{year}{2008}).

\bibitem[{\citenamefont{Pearson et~al.}(2009)\citenamefont{Pearson, Crittenden,
  and Gill}}]{PearsonJCP2009}
\bibinfo{author}{\bibfnamefont{J.~K.} \bibnamefont{Pearson}},
  \bibinfo{author}{\bibfnamefont{D.~L.} \bibnamefont{Crittenden}},
  \bibnamefont{and} \bibinfo{author}{\bibfnamefont{P.~M.~W.}
  \bibnamefont{Gill}}, \bibinfo{journal}{J. Chem. Phys.}
  \textbf{\bibinfo{volume}{130}}, \bibinfo{pages}{164110}
  (\bibinfo{year}{2009}).

\end{thebibliography}
\end{document}